\documentclass[
    ,final            
  ]
  {aipproc}

\layoutstyle{6x9}

\begin{document}

\title{Pulsar Timing with the \textit{Fermi} LAT}


\author{Paul S. Ray}{
  address={Naval Research Laboratory, 4555 Overlook Ave., SW, Washington, DC 20375, USA},
  email={paul.ray@nrl.navy.mil}
}

\author{Matthew Kerr}{
  address={Stanford University, Stanford, CA 94305, USA}
}

\author{Damien Parent}{
  address={George Mason University, Fairfax, VA 22030, resident at Naval Research Laboratory, Washington, DC 20375, USA}
}

\author{the Fermi PSC}{
  address={\textit{Fermi} Pulsar Search Consortium}
}

\begin{abstract}
We present an overview of precise pulsar timing using data from the Large Area Telescope (LAT) on \textit{Fermi}. We describe the analysis techniques including a maximum likelihood method for determining pulse times of arrival from unbinned photon data.  In addition to determining the spindown behavior of the pulsars and detecting glitches and timing noise, such timing analyses allow the precise determination of the pulsar position, thus enabling detailed multiwavelength follow up.
\end{abstract}

\maketitle


\section{Introduction}

The \textit{Fermi} Large Area Telescope (LAT) has proven to be an exceptionally powerful instrument for the study of gamma-ray pulsars (e.g. Romani et al. in this volume). Because of the large effective area, excellent background rejection made possible by the fine point spread function (PSF), precise absolute time tagging of events, and the sky survey mode of operation, the LAT is able to do precise timing of pulsars in the gamma-ray band in a way never before possible.  The continuous sky survey enables evenly sampled timing points for every pulsar in the sky over the full mission lifetime.

Exploiting the LAT as a pulsar timing instrument is a requirement for the large number of pulsars that have been found in gamma-ray blind searches.  Of the 26 known (Geminga plus the 25 discovered using the LAT; see Saz Parkinson, this volume), only 3 have been observed to pulse in radio wavelengths, and only one of those (PSR J2021+4127) is bright enough for routine radio timing. In addition, there are other pulsars that are more suitable for timing with the LAT than with radio observations, such as the very faint young pulsar PSR J1124$-$5916. Lastly, some very bright gamma-ray pulsars, such as the Crab and Vela, are good targets for LAT timing because they can be timed precisely in the gamma-ray band without any concern from time-variable dispersion measure or other propagation effects that can afflict the radio observations.

Pulsar timing allows us to characterize the rotational parameters of the pulsar and study the effects of timing noise and glitches, but for the newly-discovered gamma-ray selected pulsars perhaps the most important measurement enabled by LAT timing is precise position determination.  These timing positions achieve arcsecond accuracy, much better than the several arcminute accuracies that come from LAT photon direction measurements, and these enable deep searches for multiwavelength counterparts.

\section{Methods}

The traditional method of pulsar timing relies on radio observations taken from ground-based telescopes that are at fixed locations on the Earth.  An individual observation (typically of minutes to hours duration) is reduced to a measurement of a pulse time of arrival (TOA) referenced to an observatory clock that can be traced back to UTC. These measured `topocentric' TOAs are then fit to a pulsar timing model using software such as \textsc{Tempo2} \citep{hem06}. This code translates the TOAs to the Solar System Barycenter (SSB) using the known geodetic position of the observatory and the assumed position of the pulsar, which are parameters being fit. 

For an observatory in orbit, like the LAT, the assumption of the measurement being made at a fixed location is broken and a different technique must be used.  In order to take advantage of the established techniques and software as much as possible, we translate each photon event time to a fictitious observatory at the geocenter using the \textit{Fermi} Science Tool \texttt{gtbary} with the \texttt{tcorrect=geo} option. The geocentric time is the satellite time corrected for geometric light travel time to the geocenter. It does not include relativistic terms in the correction. The geocentric photon time $t_\mathrm{geo}$ is defined as
\begin{equation}
t_\mathrm{geo} = t_\mathrm{obs} + \frac{\mathbf{r}_\mathrm{sat}}{c} \cdot \mathbf{\hat{n}}_\mathrm{psr},
\end{equation}
where $\mathbf{r}_\mathrm{sat}$ is the vector pointing from the geocenter to the spacecraft, $\mathbf{\hat{n}}_\mathrm{psr}$ is a unit vector pointing in the direction of the pulsar (here assumed to be at an infinite distance), and $c$ is the speed of light.
 We then use \textsc{Tempo2} to generate phase prediction files (polycos) for an observatory at the geocenter and we use those polycos to compute a pulse phase for each photon. Finally, we subdivide the data set into $N_\mathrm{TOA}$ equally-spaced intervals and compute a mean TOA for each interval.

\begin{figure}
\includegraphics[width=3.0in]{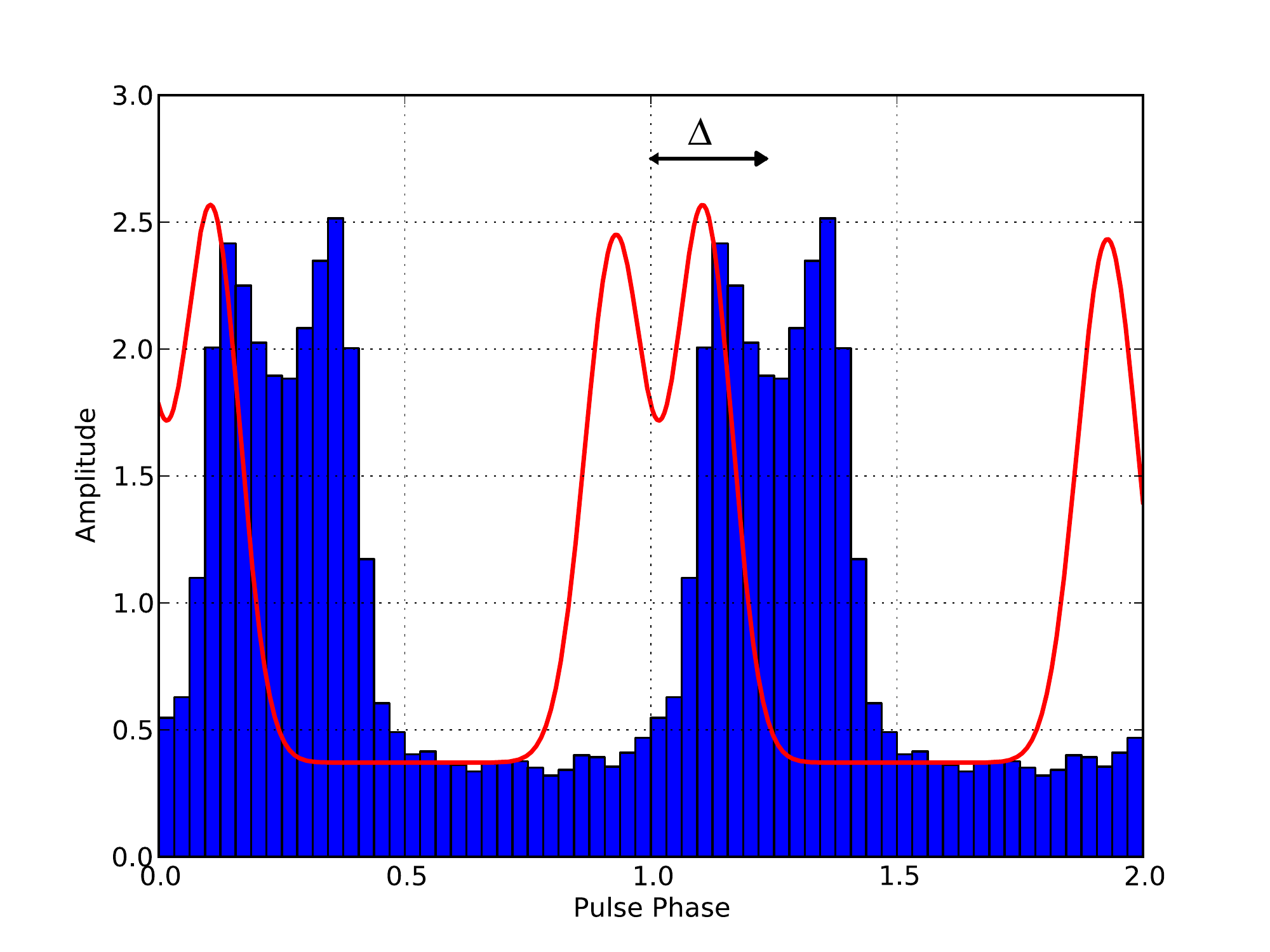}
\caption{Illustration (from \citep{rkp+10}) of the measurement of the offset between a measured pulsar light curve (blue histogram) and a template profile (red curve). The TOA is then determined by adding the measured $\Delta$ to the known observation start time.\label{lcfig}}
\end{figure}

The traditional method for computing a TOA would be to bin the photons into a phase histogram and determine the offset $\Delta$ (see Figure \ref{lcfig}) by cross-correlating with a high signal-to-noise template having the same binning. However, since (even with the LAT) gamma-ray data are very sparse we often make TOA measurements from as few as 50--100 photons. In this case it is preferable to use an unbinned maximum likelihood method to determine the TOA from the set of measured photon phases themselves. Here we treat the template profile as a periodic probability density function and compute the likelihood that the observed collection of photon phases arose from that distribution.  The log likelihood is then maximized over the phase shift $\Delta$ to obtain the TOA.  This requires a continuous function representation of the template profile. We use three functions in this work: multi-gaussian functions, kernel density estimators (KDE) with phase-dependent bandwidth, and empirical Fourier decompositions. The preferred functional form depends on the pulse shape and signal-to-noise ratio of the pulsar and the parameters of the template are determined from a maximum likelihood fit to the full mission dataset. The method is described in more detail by \citet{rkp+10}. The TOAs determined in this way can be fit with \textsc{Tempo2} using the observatory code `coe' for the center of the Earth. The full flexibility of \textsc{Tempo2} to fit for astrometric and spin parameters are thus available and the LAT TOAs can be combined with radio TOAs easily.

\section{Results}

These LAT pulsar timing techniques have been used for numerous published results on the 24 gamma-ray selected pulsars discovered in blind searches \citep{rkp+10,BSP2}, the LAT study of Geminga \citep{LATGeminga}, a detailed LAT study of the Vela pulsar \citep{Vela2}, and for refining the timing models for three new millisecond pulsars discovered in radio searches of LAT unassociated sources \citep{rrc+10}. Here we summarize a few of these results to demonstrate the power of LAT pulsar timing.

\begin{figure}
\includegraphics[width=3.0in]{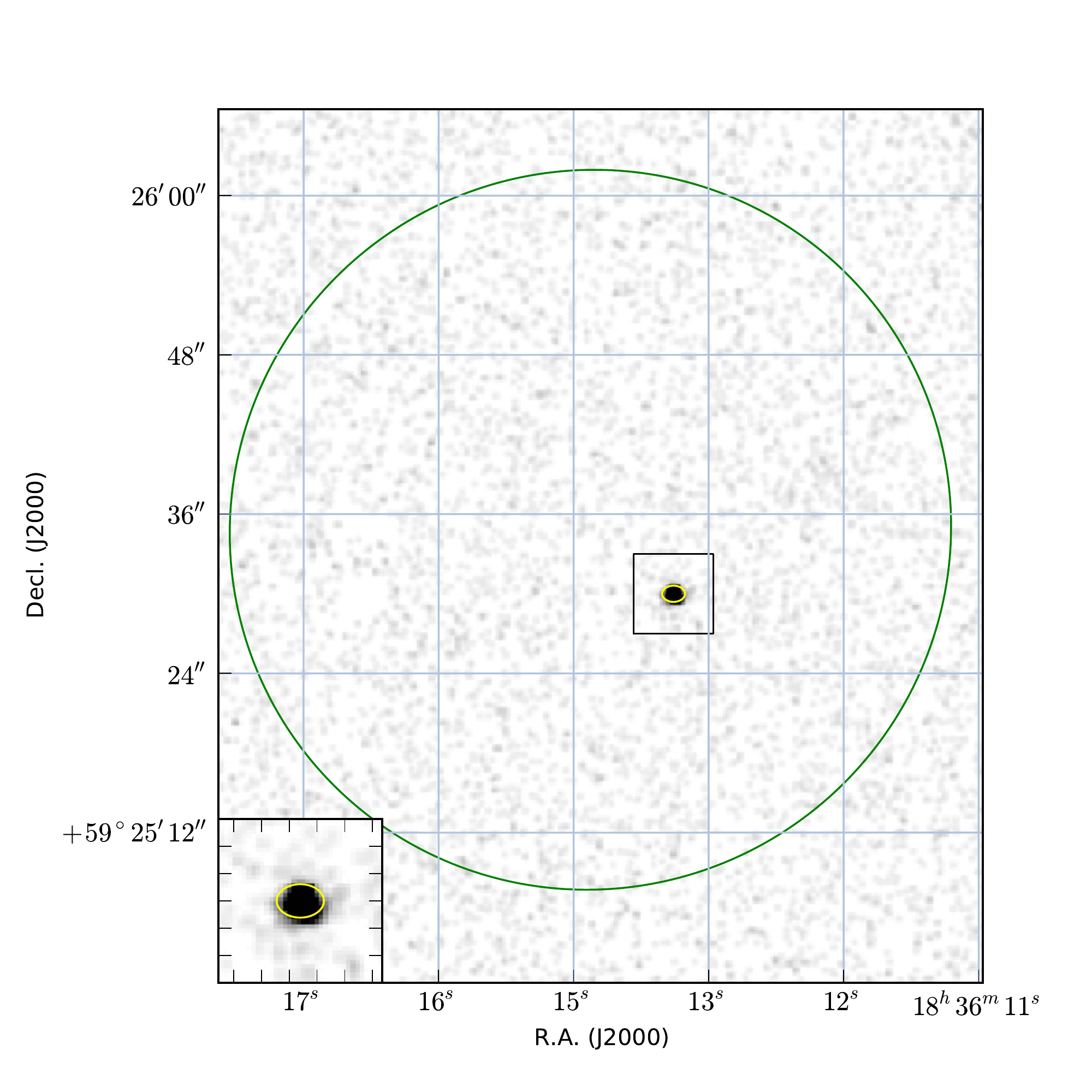}
\includegraphics[width=3.0in]{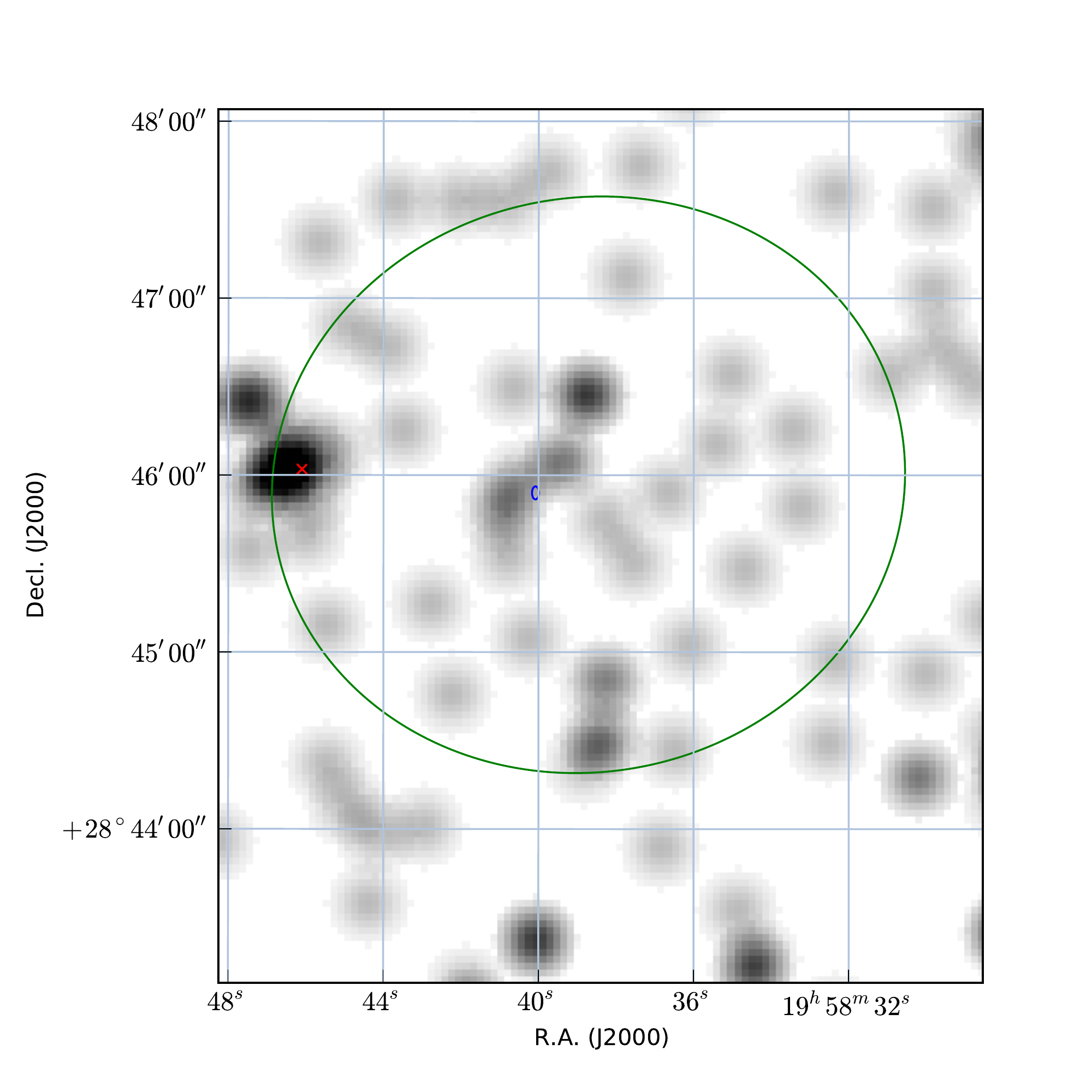}
\caption{Two examples from \citep{rkp+10} of position determinations from LAT timing. The left panel shows a Chandra X-ray image of PSR J1836+5925. The large green ellipse is the 95\% confidence position determination from LAT photon analysis, while the small yellow ellipse is the timing position (shown in more detail in the inset). The timing position clearly confirms the association with the X-ray source proposed as the counterpart before the \textit{Fermi} launch (see \citep{hcg07}). The right panel shows a Swift X-ray image of PSR J1958+2846. Here the timing position (small blue ellipse) argues against the X-ray source (marked with a red X) proposed as a possible counterpart of the pulsar in the discovery paper \citep{LATBlindSearch}. \label{posfig}}
\end{figure}

In Figure \ref{posfig}, we show two examples of position determinations based on LAT pulsar timing. In one case, the proposed X-ray counterpart is strikingly confirmed, while in the other the timing does not support the initially proposed X-ray association. Other cases are less clear, because of the influence of timing noise on the position determination.  Timing noise is a stochastic process that can induce significant red noise in the residuals to a simple timing model \citep{hlk+04}. While these residuals can be accounted for by adding higher order frequency derivatives or sinusoidal terms to the timing model, there will always be a high degree of covariance between those terms and the timing position, particularly for the limited span of timing data available this early in the \textit{Fermi} mission. This covariance means that the statistical errors on the timing position are significant underestimates of the true uncertainty on the position, which must be estimated using Monte Carlo or other methods that take into account this red noise.

In addition to timing noise, the young energetic population of pulsars that the LAT detects are also prone to frequent glitches \citep{lss00}. In our timing, we have discovered glitches in several pulsars. The radio quiet PSR J0007+7303 in the supernova remnant CTA 1 exhibited a glitch of magnitude $\Delta\nu/\nu = 5.5 \times 10^{-7}$ on 2009 May 1, and the very radio faint PSR J1124$-$5916 had a glitch of magnitude $\Delta\nu/\nu = 1.6 \times 10^{-8}$ on 2009 December 26. Neither of these pulsars can be timed in the radio band and so the LAT timing is the only way to study the glitch behavior of these systems.

\begin{figure}
\includegraphics[width=3.0in]{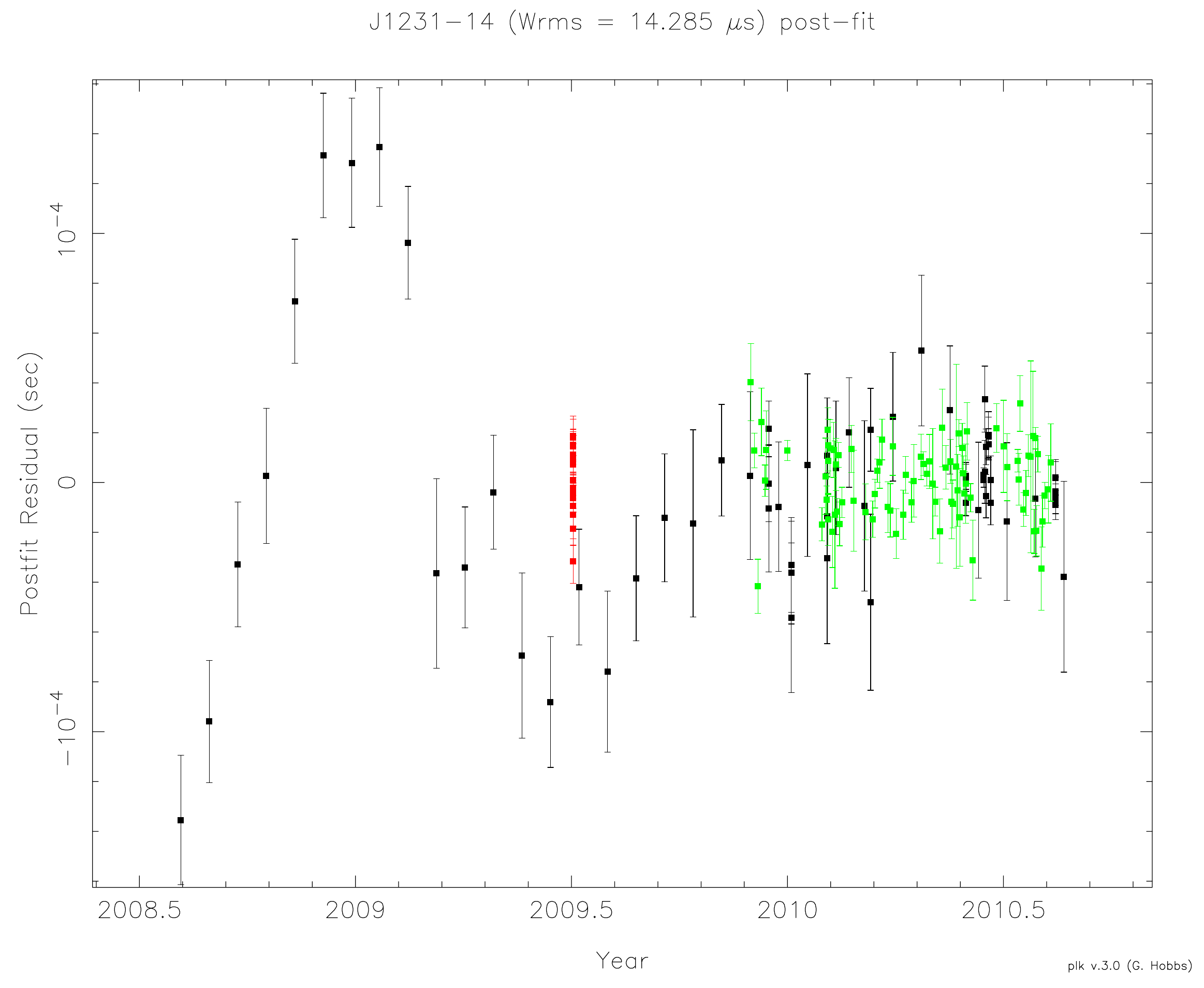}
\includegraphics[width=3.0in]{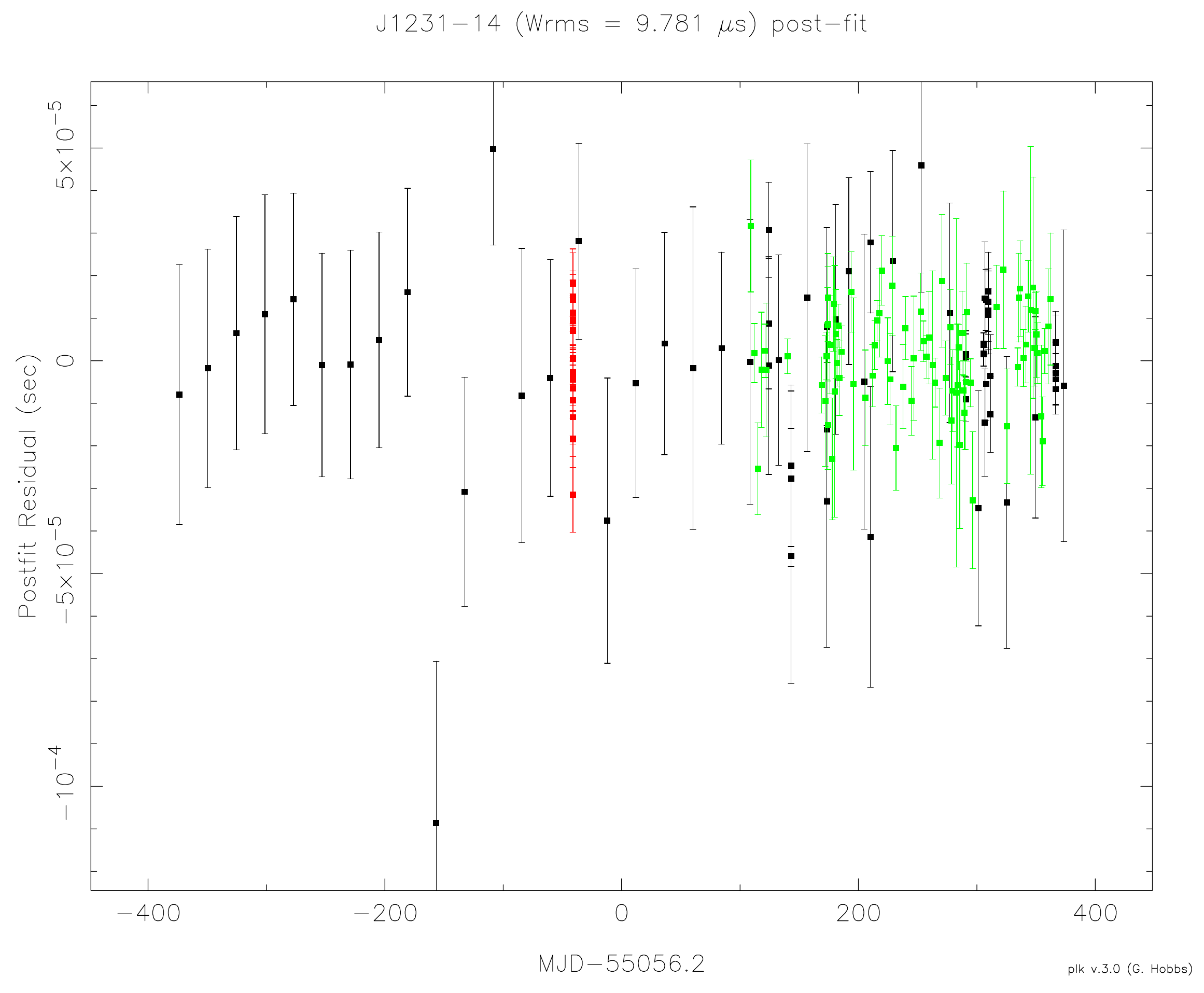}
\caption{The left panel shows the combined timing residuals with both radio and LAT TOAs for PSR J1231$-$1411 with the best fit model where the position is held fixed. The right panel shows the residuals obtained after adding proper motion to the fit.\label{fig:pm}}
\end{figure}

Besides just the young gamma-ray selected pulsars, these methods have been highly productive when applied to millisecond pulsar timing as well, achieving accuracies of tens of microseconds on the TOA measurements for the brightest gamma-ray MSPs.  Recent radio searches have revealed a large population of MSPs among the LAT unassociated source population (see Hessels et al., this proceedings). Since these are new discoveries, the baselines for radio timing is very small. In several cases, we have been able to combine the radio timing data with LAT timing to extend the timing solutions back to the beginning of the \textit{Fermi} mission.  This has allowed us in several cases to improve the position determination and make a precise measurement of the frequency derivative that was not possible with the radio data alone. In most cases, the orbital parameters of these MSPs must be determined solely from the radio data because the orbits are much shorter than the time it takes to make a significant detection of the LAT pulsations. However, in one case, PSR J0614$-$3329 \citep{rrc+10}, the orbital period is 53.6 days and we were able to generate LAT TOAs with a spacing of about 2 weeks and thus contribute directly to the determination of the orbit in this system.  In one other case, PSR J1231$-$1411, there are significant residuals observed when combining the LAT TOAs with the radio data, see Figure \ref{fig:pm}. The simplest model that accounts for these observations includes a significant proper motion of 120 $\pm$ 11 mas/yr.  This large a proper motion would have significant implications for the measured parameters of the system, and whether this is the correct model will have to be determined by further long-term timing of this pulsar, but this clearly illustrates the power of pulsar timing with the LAT.

\section{Summary}

As we have shown, the \textit{Fermi} LAT is a powerful instrument for gamma-ray pulsar timing. LAT timing is an important tool for studying a range of pulsars from the very brightest to those that have yet to be detected in any other wave band. LAT TOAs achieve sub-millisecond accuracy on many pulsars and tens of microsecond accuracy on bright millisecond pulsars and the very bright Crab and Vela pulsars.  LAT timing models have been determined for all of the pulsars discovered in blind searches of the LAT data as well as a number of pulsars that are better timed with the LAT. An important result of the timing analysis is the determination of precise (arcsecond) positions for these pulsars that enable multiwavelength follow up in the radio, optical, and X-ray bands. Timing parameters for LAT pulsars are made available via the \textit{Fermi} Science Support Center\footnote{\url{http://fermi.gsfc.nasa.gov/ssc/data/access/lat/ephems/}}.


\begin{theacknowledgments}

The \textit{Fermi} LAT Collaboration acknowledges support from a number of agencies and institutes for both development and the operation of the LAT as well as scientific data analysis. These include NASA and DOE in the United States, CEA/Irfu and IN2P3/CNRS in France, ASI and INFN in Italy, MEXT, KEK, and JAXA in Japan, and the K.~A.~Wallenberg Foundation, the Swedish Research Council and the National Space Board in Sweden. Additional support from INAF in Italy and CNES in France for science analysis during the operations phase is also gratefully acknowledged.

\end{theacknowledgments}



\bibliographystyle{aipproc}   

\bibliography{psrtiming}

\begin{thebibliography}{10}
\expandafter\ifx\csname natexlab\endcsname\relax\def\natexlab#1{#1}\fi
\providecommand{\enquote}[1]{``#1''}
\expandafter\ifx\csname url\endcsname\relax
  \def\url#1{\texttt{#1}}\fi
\expandafter\ifx\csname urlprefix\endcsname\relax\def\urlprefix{URL }\fi
\providecommand{\eprint}[2][]{\url{#2}}

\bibitem[{Hobbs} et~al.(2006)]{hem06}
G.~B. {Hobbs}, R.~T. {Edwards}, and R.~N. {Manchester}, \emph{MNRAS}
  \textbf{369}, 655 (2006).

\bibitem[{Ray} et~al.(2010)]{rkp+10}
P.~S. {Ray}, M.~{Kerr}, D.~{Parent}, et~al., \emph{ApJ}  (2010), submitted,
  arXiv:1011.2468.

\bibitem[{Saz Parkinson} et~al.(2010)]{BSP2}
P.~M. {Saz Parkinson}, et~al., \emph{ApJ} \textbf{725}, 571--584 (2010).

\bibitem[{Abdo} et~al.(2010{\natexlab{a}})]{LATGeminga}
A.~A. {Abdo}, et~al., \emph{\apj} \textbf{720}, 272--283 (2010{\natexlab{a}}),
  (Geminga).

\bibitem[{Abdo} et~al.(2010{\natexlab{b}})]{Vela2}
A.~A. {Abdo}, et~al., \emph{\apj} \textbf{713}, 154--165 (2010{\natexlab{b}}),
  (Vela2).

\bibitem[{Ransom} et~al.(2010)]{rrc+10}
S.~M. {Ransom}, P.~S. {Ray}, F.~{Camilo}, et~al., \emph{ApJ}  (2010), in press,
  arXiv:1012.2862.

\bibitem[{Hobbs} et~al.(2004)]{hlk+04}
G.~{Hobbs}, A.~G. {Lyne}, M.~{Kramer}, C.~E. {Martin}, and C.~{Jordan},
  \emph{\mnras} \textbf{353}, 1311--1344 (2004).

\bibitem[{Halpern} et~al.(2007)]{hcg07}
J.~P. {Halpern}, F.~{Camilo}, and E.~V. {Gotthelf}, \emph{\apj} \textbf{668},
  1154--1157 (2007).

\bibitem[{Abdo} et~al.(2009)]{LATBlindSearch}
A.~A. {Abdo}, et~al., \emph{Science} \textbf{325}, 840 (2009), (16 Blind Search
  Pulsars).

\bibitem[{Lyne} et~al.(2000)]{lss00}
A.~G. {Lyne}, S.~L. {Shemar}, and F.~G. {Smith}, \emph{\mnras} \textbf{315},
  534--542 (2000).

\end{thebibliography}

\IfFileExists{\jobname.bbl}{}
 {\typeout{}
  \typeout{******************************************}
  \typeout{** Please run "bibtex \jobname" to optain}
  \typeout{** the bibliography and then re-run LaTeX}
  \typeout{** twice to fix the references!}
  \typeout{******************************************}
  \typeout{}
 }

\end{document}